\documentclass[11pt,twoside]{article}
\usepackage{adassconf}

\begin{document}

\paperID{O8-3}

\title{Big Science with a Small Budget: Non-Embarrassingly Parallel Applications in a
Non-Dedicated Network of Workstations}
\titlemark{Parallel Applications in a Non-Dedicated NOW}

\author{\'Angel de Vicente, Nayra Rodr\'iguez}
\affil{Instituto de Astrof\'isica de Canarias, 38200 La Laguna (Tenerife), Spain}

\contact{Angel de Vicente}
\email{angelv@iac.es}

\paindex{de Vicente, A.}
\aindex{Rodr\'iguez, N.}

\authormark{de Vicente \& Rodr\'iguez}


\begin{abstract}
Many astronomers and astrophysicists require large computing resources for their
research, which are usually obtained via dedicated (and expensive) parallel
machines. Depending on the type of the problem to be solved, an alternative
solution can be provided by creating dynamically a computer cluster out of
non-dedicated workstations using the Condor High Throughput Computing System and
the Master-Worker (MW) framework. As an example of this we show in this paper
how a radiative transfer application previously coded with MPI is solved using
this solution without the need for dedicated machines.
\end{abstract}

\section{Introduction}
A common scenario for many researchers is the need to run a parallel application
in an expensive and usually crowded dedicated machine while at the same time
dozens or hundreds of workstations sit idle nearby. If we want to make use of
these idle machines (without disrupting their regular users) to run parallel
applications we will need: 1) a system that monitors user activity in each
workstation, adding it to a dynamically built cluster only when inactive; 2) a
parallel programming solution that does not assume that machines will be
dedicated and that can easily deal with machines joining and leaving the cluster
at any time. A solution that satisfies these two issues is offered by the
combination of the
\htmladdnormallinkfoot{Condor}{http://www.cs.wisc.edu/condor/} system and the
\htmladdnormallinkfoot{MW}{http://www.cs.wisc.edu/condor/mw/} tool.

\section{Condor-MW}
Condor is a very popular
workload management system developed at the University of Wisconsin-Madison, whose main
strength is harnessing otherwise wasted CPU power from idle workstations, being
very appropriate for parametric studies and other embarrassingly parallel
applications, {\em i.e.} those applications that can be divided into tasks that don't
depend on each other. But non-embarrassingly parallel applications are often
developed using the ``de facto'' standard for message passing, MPI, which
assumes that the application will run on dedicated machines and therefore is not
well suited for Condor.

If an application can be developed following the popular master-worker paradigm,
then a suitable solution to run parallel applications within Condor is given by
MW, which ``[\dots] is a tool for making a master-worker style application that
works in the distributed, opportunistic environment of Condor''. It can use
different methods for message passing and it will automatically handle all the
complexity of machines (possibly of different architectures and/or Operating
Systems) joining in, suspending, resuming, leaving, etc.

From the implementation point of view, MW is a set of C++ abstract base classes
and in order to use MW one needs to implement three classes:
    \begin{itemize}
      \item MWDriver, the master process and the control for distributing tasks.
      \item MWTask, the code to hold the data and the results of a work unit.
      \item MWWorker, the code to initialize a worker and to execute its tasks.
    \end{itemize}

For each class only a few functions need to be implemented. For example, for
the MWDriver we would need to implement the functions get\_userinfo(),
setup\_initial\_tasks(), pack\_worker\_init\_data() and
act\_on\_completed\_task(), whose functionality should be  clear from their names.

In Figure 1 we can see the relation between MW and Condor. The MWDriver is run
in the master machine and communicates with Condor to obtain worker
machines. Once a minimum number of workers is obtained, the master will send
them tasks to be performed, and the workers will report back to the driver when
finished, perhaps obtaining new tasks to compute. The communication between the
master and the workers can be performed in a number of ways, one of them being
with PVM (as pictured in Figure 1).

\begin{figure}
\epsscale{0.6}
\plotone{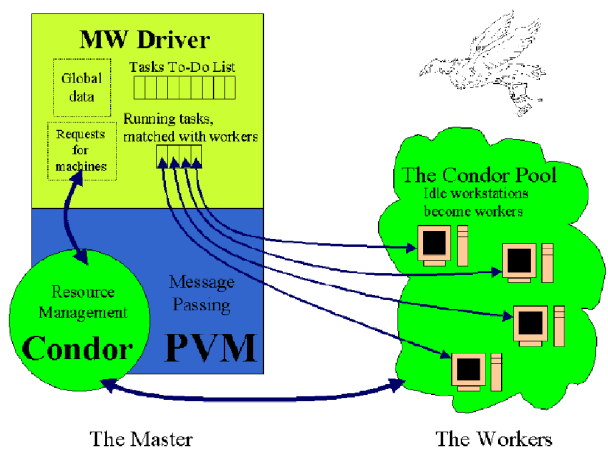}
\caption{Relationship between the MWDriver, Condor, and PVM, from (Goux,
  Linderoth \& Yoder, 2000)} 
\end{figure}

\section{Example application}

\begin{figure}
\epsscale{0.3}
\plotone{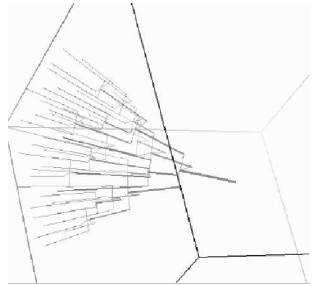}
\caption{A base ray spawns child rays, from (Abel \& Wandelt, 2002)}
\end{figure}

Using Condor-MW we implemented a radiative transfer application previously coded
using MPI in order to explore the photoionization equilibrium in non-homogeneous
HII regions. Details of the physical problem can be found in (Rodr\'iguez, 2004),
but for our purpose it will suffice to say that the program performs an adaptive
ray tracing in a three dimensional region and computes the fraction of ionized H
for each volume unit in the region. The ray tracing from the source follows the
method suggested by (Abel \& Wandelt, 2002) and is performed by the construction
of a ray tree. This tree has a ``base'' number of rays at the source, which are
divided in four new branches when necessary to assure that all the illuminated
cells are traversed by a sufficient number of rays (see Figure 2).

In the existing MPI code each ray at the base level is a task, and each worker
is responsible for the construction of its ray tree and the calculation of the
ionization changes due to its ray. Given that the MPI code had been already
implemented using the Master-Worker paradigm, it was simple to translate it into
a Condor-MW program. Nevertheless, two important changes were introduced in the
code:

\begin{itemize}
\item When using Condor, we cannot assume that a worker will be available for
  a long period. Thus, we changed the code so that a task involves the
  calculation of the ray ionization changes until it splits. Then, four new
  tasks will be created for each of the split rays. With this, the number of
  tasks increases considerably and they will not be independent any longer (the
  spectra calculated at each splitting point will be the input for the four
  newly created tasks).
\item In the MPI code the whole matrix representing the region of study was
  created for each worker, and the whole result matrix was passed back to the
  master. Since only a fraction of the matrix values were changed (those in the
  path of the given ray), it was more efficient to pass back to the master a
  matrix with only the changed values, thus decreasing significantly the amount
  of transfered data.
\end{itemize}

By changing the application in these two ways, the application can run very
efficiently in the Condor-MW environment, as can be seen in Table 1. For the
comparison we used our 32 Xeon 2.8 Ghz CPUs Beowulf-type cluster (peak
performance above 78 Gflops) and our 200 workstations (150 Linux, 50 Solaris)
Condor pool.

\begin{table}
\caption{Performance with Beowulf cluster and Condor-MW.}
\begin{tabular}{|l|l|l|}
\cline{2-3}
\multicolumn{1}{l|}{} & n=10 & n=20 \\ \hline \hline
Beowulf ({\em 8 CPUs}) & 4 hours & 38.45 hours \\ \hline
MW ({\em 8 Workers}) & 1.66 hours & \\ \hline
MW ({\em n=10, 50 Workers; n=20, 80 Workers}) & 32 minutes & 2 hours \\ \hline
\end{tabular}
\end{table}

For a small matrix of size 10, the MPI code running in 8 CPUs in the cluster
lasted 4 hours, while the Condor-MW version with only 8 workers lasted 1.66
hours. Other things being equal, we would expect the Condor-MW to be similar or
slower than the version in the Beowulf, but this gain is obviously due to the
improvements in the code, and not due to the environment itself. When requesting
50 workers, the Condor-MW version only lasted 32 minutes. For the matrix of size
20, the time needed in the Beowulf was 38.45 hours, while the same problem was
solved with the Condor-MW solution in 2 hours when requesting 80 workers.

\section{Conclusions}
As we have seen in this paper, Condor-MW can provide us with a suitable solution
for running parallel applications in a non-dedicated network of workstations if
our application can fit the Master-Worker paradigm. Even though the performance
for our example application is very good even when compared with our Beowulf
cluster, the main advantages of using the Condor-MW environment are:

\begin{itemize}
\item Existing workstations at an institution can be configured as Condor
  resources, thus allowing us to run non-embarrassingly parallel applications
  without the need for expensive investment in dedicated machines.
\item The Condor-MW code can run for months, using only a few workers when the
  Condor pool is very busy and automatically increasing the number of workers as
  these become available.
\item Worker fault-tolerance is built-in. If a worker dies in the middle of a
  computation, the master will detect its inactivity and its task will be
  assigned to another worker automatically.
\end{itemize}

\end{document}